\documentclass[prl,showpacs,twocolumn,reprint,amsmath,amssymb]{revtex4}
\usepackage{amsmath}
\usepackage{graphicx}
\usepackage{multirow}
\usepackage{dcolumn}
\usepackage{bm}
\usepackage[utf8]{inputenc}
\begin{document}
\title{Optimizing thermoelectric properties of filled \\
MPt$_4$Ge$_{12-x}$Sb$_x$ skutterudites by band engineering}
\author {M. X. Chen}
\affiliation{Institute for Physical Chemistry, University of Vienna, Sensengasse
8/7, 1090 Vienna, Austria}
\author {R. Podloucky}
\affiliation{Institute for Physical Chemistry, University of Vienna, Sensengasse
8/7, A-1090 Vienna, Austria}
\author{S. Humer}
\affiliation{Institute of Solid State Physics, Vienna University of Technology
A-1040 Vienna, Austria}
\author{E. Bauer}
\affiliation{Institute of Solid State Physics, Vienna University of Technology
A-1040 Vienna, Austria}
\author{P. Rogl}
\affiliation{Institute for Physical Chemistry, University of Vienna, Sensengasse
8/7, 1090 Vienna, Austria}

\date{\today}

\begin{abstract}
On the basis of density functional theory (DFT) calculations thermoelectric
properties are derived for Sb-doped skutterudites 
MPt$_4$Ge$_{12-x}$Sb$_x$ with M=Ba,La,Th. It is predicted that the originally
very small  absolute values of Seebeck
coefficients $|S|$ of the undoped compounds is increased by factors of 10 or
more for suitable dopings. The optimal dopings correspond  to a "magic" valence electron number for
which all electronic states up to a (pseudo)gap are filled. The theoretical
findings are corroborated by measurements of $S$ for
LaPt$_4$Ge$_{12-x}$Sb$_x$ skutterudites. 
DFT derived vibrational rattling-like modes for  LaPt$_4$Ge$_{12}$ 
indicate a small value for the lattice
thermal conductivity which in combination with a large value of $S^2$
makes the La-based skutterudites appear as promising thermoelectric materials.
\end{abstract}

\maketitle
Thermoelectric materials are of scientific as well as technological interest
in particular because of their potential for environment friendly electric power generation.
The thermoelectric performance of a material is characterized by the dimensionless figure of merit,
\begin{equation}
ZT = S^2T \sigma / (\kappa_{el}+\kappa_{ph}), 
\label{eq:merit}
\end{equation}
where $S$ represents the Seebeck coefficient, 
$\sigma$ the electrical conductivity, and $\kappa_{el}$,  $\kappa_{ph}$ are
the thermal conductivities attributed to electrons and phonons, respectively.
Skutterudites appear as promising thermoelectric materials by combining
suitable electronic, vibrational and thermal  properties. This is due to 
their crystal structure which contains large voids formed by framework atoms. 
Filling the voids by heavier elements  may reduce $\kappa_{ph}$ due to the
appearance of low-lying optical (so-called rattling)
modes~\cite{nolas_skutterudites:_1999}.
Additionally these filler atoms may provide an appropriate number of
valence electrons for making $S^2$ large.
At present, a \textit{ZT} at elevated temperatures of about 1.6
is achieved by filled pnictide skutterudites~\cite{nolas_skutterudites:_1999}.
Very recently, values of $ZT=1.7$ at 800 K were reported for the unfillled
skutterudite Fe$_{2.8}$Co$_{1.2}$Sb$_{12}$ under severe plastic
deformation~\cite{rogl_2012}. 

Ge-based skutterudites MPt$_4$Ge$_{12}$ were recently synthesized~ 
\cite{bauer_2007,bauer_2008, gumeniuk_2008,toda_2008,gumeniuk_2010,nicklas_2011}
for which the cage framework is formed entirely by Ge atoms. 
Their Seebeck coefficients are rather low as compared to
pnictogen-based skutterduites~\cite{bauer_2007,gumeniuk_2008,tran_2009}.

The aim of the present work is to utilize density functional theory results for
understanding the physical properties of Ge-based and Sb-doped filled skutterudites
in order to provide a recipe for improving their thermoelectric performance.
The quantities provided by DFT are the electronic structure, the Seebeck coefficient,
the energy of formation as a function of doping and vibrational
properties when searching for rattling modes.

Previously, efforts have been made for investigating the
electronic structure in order to optimize the thermoelectric performance.  Mahan
and Sofo~\cite{mahan_1996}  proposed that a delta-shaped transport distribution function
originating  from  very narrow electronic density of states (DOS) around the Fermi energy
$E_F$ is required to maximize ZT,  as also recently hinted by
experiment~\cite{heremans_2008}.  For this purpose, doping is essential, as it
will be applied also in the present study.  On the other hand, the required
singularity of the  DOS is not accomplishable in many real thermoelectric materials
although they possess  promising thermoelectric properties~
\cite{mishra_1997,yang_2000,zeiringer_2011,zhang_2011}. Based on a
tight-binding concept it was argued by Zhou \textit{et al.}~
\cite{zhou_2011} that extremely narrow bands are not optimal for
maximizing ZT but rather the type of scattering mechanism of the charge carriers 
is of importance.  Here, we propose a scheme of engineering the
electronic structure of MPt$_4$Ge$_{12}$ compounds by investigating the
substitutionally doped compounds  MPt$_4$Ge$_{12-x}$Sb$_x$.  In fact, we
predict an increase of the Seebeck coefficient by a factor 10 and more upon 
doping. These findings are  corroborated by new experimental
results for LaPt$_4$Ge$_{12-x}$Sb$_x$.

\begin{figure}[!h]
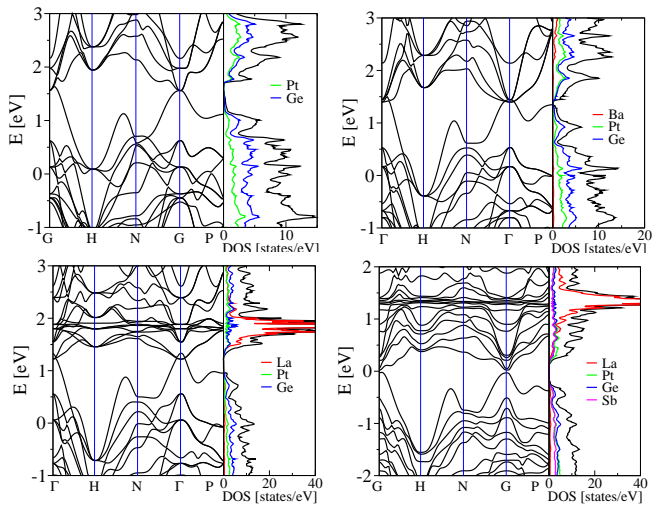

   \centering
      \includegraphics[width=0.23\textwidth]{fig2a.eps}
\hspace*{0.2cm}
      \includegraphics[width=0.23\textwidth]{fig2b.eps} \\
      \includegraphics[width=0.23\textwidth]{fig2c.eps}
\hspace*{0.2cm}
      \includegraphics[width=0.23\textwidth]{fig2d.eps} 
    \caption{Electronic structure of Pt$_4$Ge$_{12}$ (top left),
    BaPt$_4$Ge$_{12}$ (top right), LaPt$_4$Ge$_{12}$ (bottom left), 
    and LaPt$_4$Ge$_7$Sb$_5$ (bottom right). 
}
    \label{fig:dft2}
  \end{figure}

According to Boltzmann's transport theory, the Seebeck tensor is defined as 
 \begin{equation}
S(T,N) =  -\frac{1}{|e|T}K_0^{-1}K_1
  \label{eq:S1}
  \end{equation}
with the electronic charge $-|e|$.  The total number of electrons $N$ defines the chemical potential $\mu$
by
\begin{equation}
\int_{-\infty}^{\infty} n(E) f(E,\mu,T) dE = N
\label{eq:K_n}
\end{equation}
when integrating the DOS $n(E)$ weighted by the
Fermi--Dirac function $f=1/(1+\exp((E-\mu)/kT)$. In the spirit of a
rigid band model (i.e. $n(E)$ remains unchanged upon varying N) the components
of the tensor $S$  become now functions of $N$ and temperature $T$.
The tensor $K_n$  is associated with the electronic band structure by
 \begin{eqnarray}
\begin{split}
K_n =   -\frac{1}{4\pi^3}\sum_{i,\mathbf{k}} 
 \mathbf{v}_i(\mathbf{k}) \otimes \mathbf{v}_i(\mathbf{k})
\tau_i(\mathbf{k}) \times  \\
    (\varepsilon_i(\mathbf{k})-\mu)^n f'(\varepsilon_i(\mathbf{k}))
  \label{eq:S(k)}
\end{split}
  \end{eqnarray}
in which $\varepsilon_i(\mathbf{k})$ and $\mathbf{v}_i(\mathbf{k})$ are the energy
eigenvalue and the corresponding band velocity  for vector $\mathbf{k}$ and for
band $i$. The quantity
$\tau_i(\mathbf{k})$ represents the corresponding relaxation time comprising
all scattering events. The function $f'(E)=\partial{f} / \partial{E}$  is the
derivative of the Fermi--Dirac distribution. The Seebeck tensor is diagonal with all components equal if the
system has cubic symmetry or is symmetrically averaged, which we assume from
now on.  In fact, skutterudites crystallise in a cubic crystal structure.
Within Mott's approximation derived for low
temperatures~\cite{mott_1958} the Seebeck coefficient is expressed as
\begin{eqnarray}
\begin{split}
S & = - \frac{\pi^2}{3}\frac{k_B^2T}{|e|}     \left(\frac{1}{n(E)}\frac{dn(E)}{dE}
\right. \\
  & \left. +\frac{1}{v^2(E)}\frac{dv^2(E)}{dE}
    +\frac{1}{\tau(E)}\frac{d\tau(E)}{dE} \right)_{E=E_F}
  \label{eq:mott1}
 \end{split}
  \end{eqnarray}
%
relating $S$ to the DOS by 
the first term on the right side. This term becomes large for
a large slope  and a small value of $n(E_F)$  
which could be the case e.g. for doped semiconductors. 
Assuming a constant relaxation time $\tau$ (i.e. the third term is zero)
and a parabolic behaviour of the valence bands according to
$\varepsilon(\mathbf{k})=E_v - \hbar^2 k^2/2m$
and of the conduction bands according to $\varepsilon(\mathbf{k})=\hbar^2 k^2/2m - E_c$
one arrives at the relations
\begin{equation}
S_n=-|const| \frac{k_BT}{E_F - E_c}, \quad E_F > E_c
\label{eq:S_n}
\end{equation} for an n-type semiconductor
and
\begin{equation}
S_p=|const| \frac{k_BT}{E_v - E_F}, \quad E_F < E_v
\label{eq:S_p}
\end{equation} for an p-type semiconductor.
There, $E_c$ and $E_v$ are the bottom of the conduction band and the top of the valence band,
respectively. Actually, a parabolic band behavior 
is observed for MPt$_4$Ge$_{12}$ (M=Ba,La,Th) skutterudites  below the
distinctive (pseudo)gaps, as revealed by Figs.~\ref{fig:dft2}.
Maximizing $|S|$ in the spirit of Mott's approximation requires that
$E_F$ is close to the conduction or valence band edge and that the electron velocities are
large according to the second term on the right side of Eq.~\ref{eq:mott1}. 
 
For calculating the Seebeck coefficient according to
Eqs.\ref{eq:S1} and \ref{eq:S(k)} within  the constant relaxation 
time approximation a modified version of the  BoltzTrap
program~\cite{madsen_2006} is used.
Density functional theory (DFT) calculations 
were carried out by applying the
Vienna Ab initio Simulation Package (VASP)~\cite{vasp1,vasp2}.
The exchange correlation functional is approximated by
the generalized gradient approximation as parametrized by Perdew, Burke and
Ernzerhof~\cite{perdew_1996},
and the pseudopotentials are constructed by Bl\"ochl's projector
augmented wave method~\cite{paw1,paw2}.
In all calculations the
structures were fully relaxed utilizing a
$5\times5\times5$ Monkhorst and Pack $\mathbf{k}$-point grid~\cite{monkhorst_1976} ensuring
accurate results. After structural optimization 
the eigenvalues $\varepsilon_i(\mathbf{k})$ were derived on a very fine
$25\times25\times25$ $\mathbf{k}$-mesh as needed by BoltzTrap.

Substituting Ge by Sb 
leads to a 
large homogeinity region in the phase diagram, and because of
that it is unfeasible to perform fully relaxed DFT calculations for all possible structures.  
Therefore, such calculations
were only made for the most important structures (i.e. the ones with the highest number of
symmetry equivalent atomic configurations), and out of this selection the one
with lowest total energy was chosen for finally deriving the thermoelectric
properties.
  %
  \begin{figure}
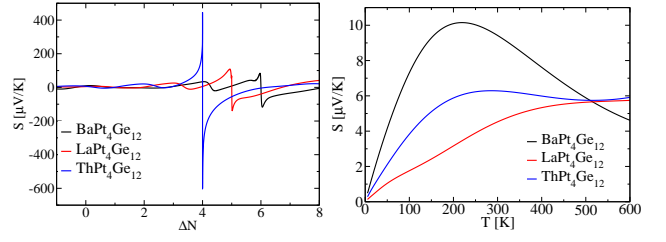

      \includegraphics[width=0.23\textwidth]{fig1a.eps} 
      \includegraphics[width=0.23\textwidth]{fig1b.eps} 
\caption{DFT derived Seebeck coefficients $S$ for MPt$_4$Ge$_{12}$,
(M=Th,La,Ba). 
Left panel: $S$ as a function of electron doping $\Delta N$ at T = 300 K;
right panel: $S$  as a function of temperature for the undoped compounds (i.e.
$\Delta N=0$).  Doping $\Delta N$ is defined
with respect to MPt$_4$Ge$_{12}$, and $\Delta N = 4,5,6$ places the
Fermi energy of MPt$_4$Ge$_{12}$ in the (pseudo)gap.
$S$ for $\Delta N \neq 0$ calculated within
the rigid band model.
}
    \label{fig:dft1}
  \end{figure}

The DFT derived Seebeck coefficients for MPt$_4$Ge$_{12}$ (M=Ba,La,Th) 
in Fig.~\ref{fig:dft1} are rather small with 
values less than 10 $\mu$V/K being consistent with experiment~\cite{bauer_2007,tran_2009}.
The left panel of Fig.~\ref{fig:dft1} showing $S$ as a function of
electron doping  reveals that $|S|$ increase strongly when the Fermi level
falls into the gap of the electronic structure, corresponding to a "magic"
doping of $\Delta N_m= 4,5,6$ for MPt$_4$Ge$_{12}$ (M=Th,La,Ba) (see also the
corresponding panels in Fig.~\ref{fig:dft2}). 
In particular 
the sign of $S$ and its behavior around the gap (i.e. change of
sign and near divergency) are reasonably described by the low-temperature
approximation for $S$ of a p- and n-conducting semiconductor according to Eqs.
\ref{eq:S_n} and \ref{eq:S_p}.
In Fig.~\ref{fig:dft2}(a) the electronic structure 
at and below the (pseudo)gap reveals a band with strong dispersion with its maximum in
$\Gamma$ (around 1.5 eV for the MPt$_4$Ge$_{12}$ compounds, 
and at $E_F$ for LaPt$_4$Sb$_5$Ge$_7$), 
which behaves rather linearly along the $\Gamma$ - H direction.
These features resemble those of CoSb$_{3}$, 
which is a narrow-gap semiconductor.
Assuming  that substitution of Co and Sb by similar
elements such as Pt and Ge does not significantly change the
(pseudo)gap formation properties a simple electron counting rule can be
constructed.  Co$_4$Sb$_{12}$ has 96 valence
electrons whereas  Pt$_4$Ge$_{12}$, BaPt$_4$Ge$_{12}$,  LaPt$_4$Ge$_{12}$
and ThPt$_4$Ge$_{12}$ have only 88, 90,91 and 92 valence electrons, respectively.
For placing now $E_F$ into the (pseudo)gap the Pt-Ge compounds have to be doped by
8,6,5 and 4 electrons ("magic" doping), correspondingly.

Figure~\ref{fig:dft2} reveals rather minor influences on the (pseudo)gap when filling the voids of
Pt$_4$Ge$_{12}$: it is preserved for Ba- and LaPt$_4$Ge$_{12}$ and becomes a
true but
small  gap for ThPt$_4$Ge$_{12}$ (not shown).  
For LaPt$_4$Ge$_{12}$ and ThPt$_4$Ge$_{12}$ bands above the conduction
band edge are rather flat due to their localized f-like character. Then,
$\frac{\partial n(E)}{\partial E}$ is becoming large and as indicated by  
Eq.~\ref{eq:mott1} $|S|$ increases accordingly. Apart from changing the sign, 
this is the reason for the asymmetry
of $S$ below and above the (pseudo)gap, with $|S|$ significantly larger
for dopings $\Delta N > \Delta N_m$.
The crucial point is to find appropriate filler atoms which provide the
correct number of doping electrons for $E_F$ falling into the
(pseudo)gap.

When the filler atom $M$ is chosen for doping, an element which provides
8 valence electrons is needed. Then, the dopant must be a lanthanide or
actinide atom with f-states as
valence states. However, due to their localized nature the f-states may strongly influence the electronic
structure at Fermi energy. For example,
a metalllic rather than semiconducting behavior is predicted for SmPt$_4$Ge$_{12}$ 
by previous calculations~\cite{gumeniuk_high-pressure_2010}.

Another way of doping and adding valence electrons could be chosen by substituting the transition element.
Noble metal elements such as Cu,Ag,Au are possible candidates providing one more electron than the
latest transition elements Ni, Pd, Pt. 
The question is, how the electronic structure is affected and if the gap feature is preserved.
When testing the replacement of Pt by Au by studying MAu$_4$Ge$_{12}$ compounds it turns out that the
lowering of the 5d-level of Au in comparison to the Pt 5d-levels  destroys the (pseudo)gap
feature and hence also the promising thermoelectrical properties.

  \begin{figure}[!hb]
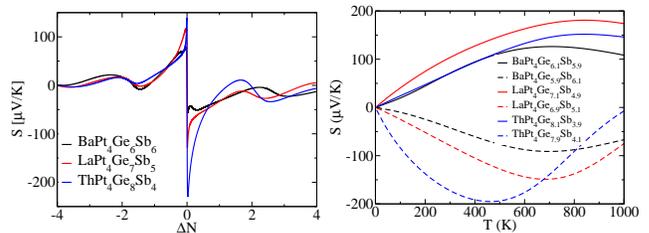

      \includegraphics[width=0.23\textwidth]{fig3a.eps} 
      \includegraphics[width=0.23\textwidth]{fig3b.eps} 
    \caption{DFT derived Seebeck coefficients as a function of doping at T=300K (left panel) and
temperature (right panel) for selected MPt$_4$Sb$_{12-x}$Ge$_x$ compounds.
The change of sign of $S$ occurs at the "magical" composition, for which the
total number of valence electrons is 96.
}
    \label{fig:dft3}
  \end{figure}

  \begin{figure}[!hb]
      \includegraphics[width=0.20\textwidth,clip]{fig4a.eps} 
 \hspace*{0.4cm}
      \includegraphics[width=0.20\textwidth,clip]{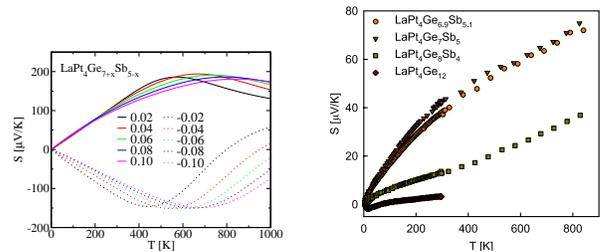} 
    \caption{DFT derived (left panel) and experimental Seebeck coefficients
(right panel) as a function of temperature  for  LaPt$_4$Ge$_{12-x}$Sb$_x$ 
compounds.
}
    \label{fig:exp1}
  \end{figure}

A third concept for substitutions  is the replacement of Ge by elements such as Sb with a larger
number of valence electrons than 4.
Filled skutterudites such as BaPt$_4$Sb$_6$Ge$_6$, LaPt$_4$Sb$_5$Ge$_7$ and ThPt$_4$Sb$_4$Ge$_8$ 
have the "magic" number of valence electrons, namely 96, and what is important
for them the desired electronic structure properties are preserved, as shown in
Fig.~\ref{fig:dft2} for the La compound as an example. Even a small gap opens
up, which is not the case for undoped LaPt$_4$Ge$_{12}$. For the
calculated thermoelectric
properties of LaPt$_4$Sb$_5$Ge$_7$  it turns out that $|S|$ is substantially
increased by more than a factor of 10 (Fig.~\ref{fig:dft3})
reaching now values even larger than  100 $\mu$V/K for the La-compound and 
more than 200 $\mu$V/K  for the Th-compound.
Figure~\ref{fig:dft3}(b) depicts temperature dependent Seebeck coefficients of 
some MPt$_4$Ge$_{12-x}$Sb$_x$ compounds around the critical composition (i.e.
"magic" doping).
Values of $|S|$ larger than 
100 $\mu$V/K are reached in a temperature range of 400-800~K depending on the
compound and its composition. From these findings one may
speculate that in general large Seebeck coefficients are obtained with a proper doping of
Sb for other Ge-based skutterudites.

Figure~\ref{fig:exp1} compares DFT derived and experimental results for
LaPt$_4$Ge$_{12-x}$Sb$_x$ compounds. Clearly, the experimental evidence
corroborates the theoretical prediction of a very strong increase of $|S|$ upon
doping by Sb. This enhancement effect seems to be more pronounced by the DFT
results, which predict a maximum of $S(T)$ ranging from 500 to 800 K for
p-type materials depending on the amount of substitutions. The DFT data also show
the sensitivity upon substitution. Increasing the amount of Sb above the critical
value (i.e. crossing the gap) changes its sign and results in comparable
(although slightly smaller) values of $|S|$. 

For maximizing the figure of merit
in Eq.~\ref{eq:merit} the power factor $S^2\sigma$ should be large. Indeed, from
the DFT calculations it was found that for LaPt$_4$Ge$_{7.1}$Sb$_{6.9}$ the
power factor reaches a maximum at 600K, at a temperature which is useful also
for technological applications. It should be noted that an absolute value of
$S^2\sigma$ cannot be obtained from the first-principles calculations, because
the relaxation time (assumed to be constant) is not known. For the Seebeck coefficient,
however, the constant relaxation time cancels out.
Nevertheless, an enhancement of at least two orders of magnitude in the power factor 
is achieved for each compound within the constant relaxation time approximation.
Experimental data for LaPt$_4$Ge$_7$Sb$_5$ at 800 K reveal a power factor of
$S^2 \rho = 7.3*10^{-5}$ W/K$^2$.

According to Eq.~\ref{eq:merit}  a large figure of merit can also be achieved
when the thermal conductivities are small. Concerning $\kappa_{el}$, the
electronic contribution, it should be proportional to the DOS at Fermi energy.
Inspecting Fig.~\ref{fig:dft2} for LaPt$_4$Ge$_7$Sb$_5$ it is obvious that the
DOS is very small just below the gap. Therefore, experimenting with the
Sb substitutions close to the critical Sb$_5$  composition might lead to a
very small value of $\kappa_{el}$.  Concomitantly the electrical
conductivity $\sigma$ becomes also small, which counteracts the enhancement
effect of $\kappa_{el}$ for the figure of merit.

Considering the phonon
contribution to the thermal conductivity the La-compound seems very promising
to yield a low value of $\kappa_{ph}$, because for LaPt$_4$Ge$_{12}$  well
localized rattling-like vibrational modes were found by DFT calculations
which resulted in a pronounced peak of La-character at 1.7 THz in the phonon
DOS.  Replacing La by Ba, which has a very similar mass but one valence electron
less, results in a very different phonon spectrum at lower frequencies without
any hint of a rattling-type feature. Results of these calculations of
vibrational properties of MPt$_4$Ge$_{12}$ compounds will be presented in a
forthcoming work~\cite{Chen_rattling_2012}.

For synthesizing the material it is important to estimate its thermodynamic
stability. Doing this by deriving the DFT enthalpy of formation $\Delta H$ it is found that
about 1 kJ/mole is gained per substituting Sb atom with the minimum of $\Delta
H$ (i.e.  strongest bonding) for the doped compound MPt$_4$Ge$_7$Sb$_5$.
Further increasing the amount of Sb 
reduces the bonding and hence the energy of formation is becoming less negative.
This behavior reflects the filling of bonding valence states upon doping
until the gap is reached for 5 Sb atoms (Fig.~\ref{fig:dft2}, right
bottom panel). Adding more Sb adds non- or antibonding conduction band states
to the total energy of the compound and hence $\Delta H$ is getting less
negative.

On the basis of DFT calculations we predict new and promising thermoelectric
materials, when in MPt$_4$Ge$_{12}$ skutterudites Ge is suitably replaced by
Sb. In particular, LaPt$_4$Ge$_{7-x}$Sb$_{5-x}$ is studied which reveals strongly
enhanced values of the Seebeck coefficient up to about 200~$\mu$ V/ K  at 600 K. 
The "magic" composition of 7 Ge and 5 Sb atoms can be understood in
terms of band engineering revealing the importance of electronic bands with
large dispersions below a (pseudo)gap, very similar to CoSb$_3$.
Samples of LaPt$_4$Ge$_{7-x}$Sb$_{5-x}$ were experimentally synthesized.
Measurements of the Seebeck coefficient strongly
supports the DFT predictions. In addition, from the DFT derived vibrational
properties of LaPt$_4$Ge$_{7}$Sb$_{5}$, which  reveal rattling modes, it can be
concluded that the phonon thermal conductivity is expected to be small, which
--in addition to the large Seebeck coefficient-- would enhance the figure of
merit.
\begin{acknowledgments}   
The authors gratefully acknowledge the support by the FWF funded project nr.
P22295-N20.  Most of the DFT calculations were done on the Vienna Scientific Cluster (VSC).
\end {acknowledgments}

\end{document}